\documentclass[
    ,final            
  ]
  {aipproc}

\layoutstyle{8x11double}

\usepackage{amsmath}
\usepackage{amsfonts}
\usepackage{amssymb}
\usepackage{graphicx}
\usepackage[english]{babel}
\usepackage{wasysym}
\usepackage{units}

\begin{document}

\newcommand\be{\begin{equation}}
\newcommand\ee{\end{equation}}
\newcommand\bea{\begin{eqnarray}}
\newcommand\eea{\end{eqnarray}}
\newcommand\bseq{\begin{subequations}} 
\newcommand\eseq{\end{subequations}}
\newcommand\bcas{\begin{cases}}
\newcommand\ecas{\end{cases}}
\newcommand{\p}{\partial}
\newcommand{\f}{\frac}

\title{Minisuperspace dynamics in a generalized uncertainty principle framework}


\author{Marco Valerio Battisti}{
address={ICRA - International Center for Relativistic Astrophysics\\
Dipartimento di Fisica (G9), Universit\`a di Roma ``La Sapienza'' P.le A. Moro 5, 00185 Rome, Italy\vspace{0.2cm}}
}

\author{Giovanni Montani}{
address={ICRA - International Center for Relativistic Astrophysics\\
Dipartimento di Fisica (G9), Universit\`a di Roma ``La Sapienza'' P.le A. Moro 5, 00185 Rome, Italy\vspace{0.2cm}},
altaddress={ENEA C.R. Frascati (Dipartimento F.P.N.), Via Enrico Fermi 45, 00044 Frascati, Rome, Italy\\ICRANET C.C. Pescara, P.le della Repubblica 10, 65100 Pescara, Italy\vspace{0.2cm}\\
{\footnotesize\ttfamily battisti@icra.it\qquad montani@icra.it}}
}

\begin{abstract} 
The minisuperspace dynamics of the Friedmann-Robertson-Walker (FRW) and of the Taub Universes in the context of a Generalized Uncertainty Principle is analyzed in detail. In particular, the motion of the wave packets is investigated and, in both the models, the classical singularity appear to be probabilistic suppressed. Moreover, the FRW wave packets approach the Planckian region in a stationary way and no evidences for a Big-Bounce, as predicted in Loop Quantum Cosmology, appear. On the other hand, the Taub wave packets provide the right behavior in predicting an isotropic Universe.  
\end{abstract}

\maketitle


\section{I. Introduction}

The so-called Generalized Uncertainty Principle (GUP) it is an immediate way to realize the old intuition about the existence of a fundamental minimal scale. In fact, a natural (Planckian) cut-off length has, in some (not yet understood) sense, to appears as soon as the smooth picture of the spacetime manifold breaks down, i.e. when the quantum effects are taken into account. 

Interest on in minimal length or GUP approach has been motivated by studies in perturbative string theory \cite{String}, considerations on the proprieties of black holes \cite{Mag} and de Sitter space \cite{Sny}. In particular, from the string theory point of view, a minimal observable length it is a consequence of the fact that strings can not probe distance below the string scale. However, in recent years, a big amount of work has been done in this active field in a wide variety of directions (see for example \cite{GUP1} and the references therein; for another application of the GUP approach to the minisuperspace dynamics, from a different point of view, see \cite{Vakili}). 

In this paper we address the question about the application of the GUP formalism in quantum cosmology. In particular, two cosmological model are discussed: i) the FRW ($k=0$) model with a massless scalar field and ii) the Taub model. For the discussion on the FRW model we refer to \cite{BM07a} and on Taub to \cite{BM07b}. 

In the first model (FRW) \cite{BM07a}, the scalar field is used as an ``relational time'' and only the scale factor is treated in the GUP formalism. As well-known in the Wheeler-DeWitt (WDW) framework the unavoidable classical singularity cannot be solved and the wave packet follow a classical trajectory up to the ``initial'' singularity \cite{Ish,APS}. In the GUP approach, as we will see, the situation is very different. In fact the wave packet, peaked at late times (at energies much smaller than the Planck's one), ``escape'' from the classical trajectory in the dynamics toward the cosmological singularity. Therefore the probability density to find the Universe near the classic time where the singularity appears goes to zero and, in some sense, our quantum Universe approach stationary states ``near the Planckian region''. In this sense the cosmological singularity is solved by the modified Heisenberg algebra.

The second model (Taub) \cite{BM07b}, is studied in the context of the ADM reduction of the dynamics. Such a representation, allows us to regard one variable, mainly the Universe volume, as a ``time'' for the dynamics. Therefore, only the physical degree of freedom of the system, which is related to the Universe anisotropy, will be treated in the GUP formalism. In the canonical case (WDW theory), the wave packets are peaked around the classical trajectories and, after the bounce on the potential wall, they fall in the cosmological singularity. On the other hand, in the GUP case, we obtain two remarkable results. i) The probability density to find the Universe is peaked ``near'' the potential wall and the wave packets show a stationary behavior. Therefore, the classically singularity will be not probabilistically privileged. ii) The value of anisotropy for which the probability amplitude is peaked corresponds to a quasi-isotropic Universe. 

The paper is organized as follows. In Sec. II the GUP framework is reviewed and Sec. III is devoted to discuss the application of this approach in quantum cosmology. In Sec. IV and V the FRW model is studied in the WDW and GUP approach, respectively and a comparison between these results is given. Finally, in Sec. VI, VII and VIII the Taub model is presented and analyzed in the WDW and in the GUP framework, respectively. Concluding remarks follow. 
  
\section{II. Quantum mechanics in the GUP framework}

In this section we briefly review some aspects and results of a non-relativistic quantum mechanics with nonzero minimal uncertainties in position \cite{Kem}. In one dimension, we consider the Heisenberg algebra generated by $\bf q$ and $\bf p$ obeying the commutation relation (in $\hbar=c=1$ units)  
\be\label{modal}
[{\bf q},{\bf p}]=i(1+\beta{\bf {p}}^2), 
\ee
where $\beta$ is a ``deformation'' parameter. This commutation relation leads to the uncertainty relation
\be\label{gup}
\Delta q \Delta p\geq \f 1 2\left(1+\beta (\Delta p)^2+\beta \langle{\bf p}\rangle^2\right),
\ee
which appears in perturbative string theory \cite{String}. The canonical Heisenberg algebra can be recovered in the limit $\beta=0$ and the generalization to more dimension is straightforward, leading naturally to a ``noncommutative geometry'' for the space coordinates.

It is immediate to verify that such a Generalized Uncertainty Principle (\ref{gup}) implies a finite minimal uncertainty in position $\Delta q_{min}=\sqrt\beta$. As well-known, the existence of a nonzero uncertainty in position implies that there cannot by any physical state which is a position eigenstate. In fact, an eigenstate of an observable necessarily has vanishing uncertainty on it. To be more precise, let us assume the commutation relations to be represented on some dense domain $D\subset H$ in a Hilbert space $H$. In the canonical case, a sequence $\vert\psi_n\rangle\in D$ exists, with position uncertainties decreasing to zero. On the other hand, in presence of a minimal uncertainty $\Delta q_{min}\geq0$, it is not possible any more to find some $\vert\psi_n\rangle\in D$ such that
\be
\lim_{n\rightarrow\infty}\left(\Delta q_{min}\right)_{\vert\psi_n\rangle}=\lim_{n\rightarrow\infty}\langle\psi\vert({\bf q}-\langle\psi\vert{\bf q}\vert\psi\rangle)^2\vert\psi\rangle=0.
\ee 
Although it is possible to construct position eigenvectors, they are only formal eigenvectors and not physical states. Let us now stress that this feature comes out from the corrections to the canonical commutation relations and, in general, a non-commutativity of the ${\bf q}_i$ will not imply a finite minimal uncertainty $\Delta q_{min}\geq0$. Therefore, in the GUP approach, we can not work in the configuration space and some notion of position will recovered in the next. 

The Heisenberg algebra (\ref{modal}) can be represented in the momentum space, where the $\bf q$, $\bf p$ operators act as 
\be\label{rep}
{\bf p}\psi(p)=p\psi(p), \qquad {\bf q}\psi(p)=i(1+\beta p^2)\p_p\psi(p),
\ee
on a dense domain $S$ of smooth functions. To recover information on positions we have to study the states that realize the maximally-allowed localization. Such states $\vert\psi^{ml}_{\zeta}\rangle$ of maximal localization, which are proper physical states around a position $\zeta$, have the proprieties $\langle\psi^{ml}_{\zeta}\vert {\bf q}\vert\psi^{ml}_{\zeta}\rangle=\zeta$ and $(\Delta q)_{\vert\psi^{ml}_{\zeta}\rangle}=\Delta q_{min}$. These states are called of maximal localization, because they obey the minimal uncertainty condition $\Delta q\Delta p=\vert\langle [{\bf q},{\bf p}]\rangle\vert/2$ and therefore the following equation holds
\be
\left({\bf q}-\langle{\bf q}\rangle + \f{\langle [{\bf q},{\bf p}]\rangle}{2(\Delta p)^2}({\bf p}-\langle{\bf p}\rangle)\right){\vert\psi^{ml}_{\zeta}\rangle}=0,
\ee
which admit, in the momentum space, the following solution\footnote{The absolutely minimal uncertainty in position $\Delta q_{min}=\sqrt\beta$ and thus also the maximal localization states, are obtained for $\langle{\bf p}\rangle=0$.}
\be
\psi^{ml}_{\zeta}(p)\sim\f 1 {(1+ \beta p^2)^{1/2}} \exp\left(-i\f{\zeta}{\sqrt{\beta}} \tan^{-1}(\sqrt{\beta}p)\right),
\ee
where with $\sim$ we omit the normalization constant. As we can easily see, these states in the $\beta=0$ limit reduce to ordinary plane waves. As last step, we can project an arbitrary state $\vert\psi\rangle$ on the maximally localized states $\vert\psi^{ml}_{\zeta}\rangle$ in order to obtain the probability amplitude for a particle being maximally localized around the position $\zeta$ (i.e. with standard deviation $\Delta q_{min}$). We call these projections the ``quasiposition wave function'' $\psi(\zeta)\equiv\langle\psi^{ml}_{\zeta}\vert\psi\rangle$; explicitly, we have
\be\label{qwf} 
\psi(\zeta)\sim\int^{+\infty}_{-\infty}\f{dp}{(1+\beta p^2)^{3/2}} \exp\left(i\f{\zeta}{\sqrt{\beta}} \tan^{-1}(\sqrt{\beta}p)\right)\psi(p).
\ee
This is nothing but a generalized Fourier transformation, where in the $\beta=0$ limit the ordinary position wave function $\psi(\zeta) = \langle\zeta\vert\psi\rangle$ is recovered.

\section{III. On the GUP in the minisuperspace dynamics}

Let us discuss some aspects regarding the application of the GUP framework in quantum cosmology, i.e. in the context of a minisuperspace reduction of the dynamics. In fact, in such a theory, only a finite number of the gravitational degrees of freedom are invoked at quantum level and the remainder are set to zero by the imposition of symmetries on the spatial metric. In particular by requiring the spatial homogeneity, the (gravitational) system is described by three degrees of freedom, i.e. the three scalar factors of the Bianchi models. On the other hand, by imposing also the isotropy, we deal with a one-dimensional mechanical system, i.e. the FRW models.

Therefore, quantum cosmology is a quantum mechanical toy model (finite degrees of freedom) which is a simple arena to test ideas and constructions which can be introduced in the (not yet found) quantum general relativity. However, since at classical level the Universe dynamics is described by such symmetric models, the quantization of these seems to be necessary to answer to the fundamental questions like the fate of the classical singularity, the inflationary expansion and the chaotic behavior of the Universe toward the singularity.

In this respect, the GUP approach to quantum cosmology appears physically grounded. In fact, a generalized uncertainty principle can be immediately reproduced deforming the canonical Heisenberg algebra. In other words, the GUP scheme relies on a modification of the canonical quantization prescriptions and, in this respect, it can be reliably applied to any dynamical system. Although such a deformed commutation relation, differently from the GUP itself, has not been so far derived directly from string theory, it is one possible way in which certain features of a more fundamental theory may manifest themselves in quantizing a cosmological model.

\section{IV. The FRW model in the WDW approach}

The canonical quantization (in the sense of the WDW theory) of the homogeneous, isotropic, flat ($k=0$) cosmological model with a massless scalar field is reviewed (for more details see \cite{Ish,APS}). The Hamiltonian constraint for this model has the form
\be\label{con}
H_{grav}+H_{\phi}\equiv-9\kappa p_x^2x+\f3 {8\pi}\f{p_{\phi}^2}{x}\approx0 \quad x\equiv a^3,
\ee      
where $\kappa=8\pi G\equiv8\pi l_P^2$ is the Einstein constant and $a$ is the scale factor. In the classical theory, the phase space is $4$-dimensional, with coordinates $(x,p_x;\phi,p_{\phi})$. At $x=0$ the physical volume of the Universe goes to zero and the singularity appears. Since $\phi$ does not enter the expression of the constraint, $p_{\phi}$ is a constant of motion and therefore each classical trajectory can be specified in the $(x,\phi)$-plane. Thus $\phi$ can be considered as a relational time and the dynamical trajectory reads as
\be\label{clastra}
\phi=\pm\f 1 {\sqrt{24\pi\kappa}}\ln\left|\f x {x_0}\right|+\phi_0,
\ee
where $x_0$ and $\phi_0$ are integration constants. In this equation, the plus sign describes an expanding Universe from the Big-Bang, while the minus sign a contracting one into the Big-Crunch. We now stress that the classical cosmological singularity is reached at $\phi=\pm\infty$ and every classical solution, in this model, reaches the singularity. 

At quantum level the Wheeler-DeWitt equation, associated to the constraint (\ref{con}), tells us how the wave function $\Psi(x,\phi)$ evolves as $\phi$ changes; in this respect we can regard the argument $\phi$ of $\Psi(x,\phi)$ as an ``emergent time'' and the scale factor as the real physical variable. In order to have an explicit Hilbert space, we perform the natural decomposition of the solution into positive and negative frequency parts. Therefore, the solution of this Wheeler-DeWitt equation has the very well-known form
\be\label{solcan} 
\Psi_\epsilon(x,\phi)=x^{-1/2}\left(Ax^{-i\gamma}+Bx^{i\gamma}\right)e^{i \sqrt{24\pi\kappa}\epsilon\phi}, 
\ee
where $\gamma=\f 12(4\epsilon^2-1)^{1/2}\geq0$ and $\epsilon^2$ being the eigenvalue of the operator $\Xi/24\pi\kappa$ defined below. Thus the spectrum is purely continuous and covers the interval $(\sqrt{3}/2l_P,\infty)$ \cite{Ish}. The wave function $\Psi_\epsilon(x,\phi)$ is of positive frequency with respect to the internal time $\phi$ and satisfies the positive frequency (square root) of the quantum constraint (\ref{con}); we deal with a Sch\"odinger-like equation $-i\p_\phi\Psi=\sqrt{\Xi}\Psi$, where $\Xi\equiv24\pi\kappa\hat{x}\hat{p_x}^2\hat{x}$. 

In order to examine the behavior of the classical singularity at quantum level we have to clarify a general criteria for determining whether the quantized models actually collapse \cite{Got}. Unfortunately there is not such a rigorous criteria yet. An early idea was to impose the condition that the wave function vanishes at the singularity $a=0$ \cite{DW}, but this boundary conditions has little to do with the quantum singularity avoidance. It seems better to study the expectation values of observables which classically vanish at the singularity. In fact, $\vert\Psi(a=0,t)\vert^2$ is merely a probability density and thus, for example, one might have an evolving state that ``bounces'' (i.e. a nonsingular wave packet), even if $\vert\Psi(a=0,t)\vert\neq0$ for all $t$. On the other hand, if one could find a wave packet so that the probability $P_{\delta}\equiv\int_0^\delta\vert\Psi(a,t)\vert^2da\simeq0$ for $\delta$ being a very small quantity, then one could reasonably claim to have a nocollapse situation.

Let's now come back to the canonical FRW model. It is not difficult to see that, in this framework, the unavoidable classical singularity is not tamed by quantum effects. In fact, if one starts with a state localized at some initial time, then its peak moves along the classical trajectory and falls into the classical singularity. Additionally, from the eigenfunctions (\ref{solcan}) it is clear that the probability defined above diverges indicating that the Wheeler-DeWitt formalism does not solve the classical singularity.

\section{V. The FRW model in the GUP approach}

In this section, we analyze the quantization of the FRW ($k=0$) model in the framework of minimal length uncertainty relation \cite{BM07a}. As in the canonical case, let us decompose the solution of the ``generalized Wheeler-DeWitt equation'' into positive and negative frequency parts and focus on the positive frequency sector. Remembering that we have to work in the momentum space, the wave function reads as: $\overline{\Psi}(p,\phi)=\Psi(p)e^{i \sqrt{24\pi\kappa}\epsilon\phi}$, where from now on $p\equiv p_x$. Thus, as soon as we regard the scalar field as an ``emergent time'' for the quantum evolution, then we treat in the ``generalized'' way only the real degree of freedom of the problem: the isotropic volume $x$. Therefore the quantum equation relative to the Hamiltonian constraint (\ref{con}), considering the representation (\ref{rep}), is the following
\be\label{emu}
\mu^2(1+\mu^2)^2\f{d^2\Psi}{d\mu^2}+2\mu(1+\mu^2)(1+2\mu^2)\f{d\Psi}{d\mu}+\epsilon^2\Psi=0,
\ee  
where we have defined a dimensionless parameter $\mu\equiv\sqrt\beta p$. In order to integrate the above equation we introduce the variable $\rho\equiv\tan^{-1}\mu$, which maps the region $0<\mu<\infty$ to $0<\rho<\pi/2$. Then, performing another change of variables: $\xi\equiv\ln(\sin\rho)$ ($-\infty<\xi<0$), equation (\ref{emu}) reduces to
\be
\f{d^2\Psi}{d\xi^2}+2\left(\f{1+e^{2\xi}}{1-e^{2\xi}}\right)\f{d\Psi}{d\xi}+\epsilon^2\Psi=0,
\ee 
which can be explicitly solved and whose general solution reads as
\be
\Psi_\epsilon(\xi)=C_1 e^{-\xi(1+\alpha)}\left(1+b_+e^{2\xi}\right)+C_2 e^{-\xi(1-\alpha)}\left(1+b_-e^{2\xi}\right),
\ee
where $\alpha=\sqrt{1-\epsilon^2}$ and $b_\pm=1\pm\alpha/(1\mp\alpha)$. At this point we have to analyze the ``quasiposition wave function'' relative to this problem in order to make a first comparison with the canonical case, in particular with the wave function (\ref{solcan}). In agreement with formula (\ref{qwf}) we have
\begin{multline}\label{quapos}
\Psi_\epsilon(\zeta)=\int_{-\infty}^0d\xi \exp{\left(\xi+i\zeta\tan^{-1}\left(\f{e^\xi}{\sqrt{1-e^{2\xi}}}\right)\right)}\\
\times\left[C_1 e^{-\xi(1+\alpha)}\left(1+b_+e^{2\xi}\right)+C_2 e^{-\xi(1-\alpha)}\left(1+b_-e^{2\xi}\right)\right],
\end{multline}
where $\zeta$, in this case, is expressed in units of $\sqrt\beta$. Thus we can easily see that our ``quasiposition wave function'', i.e. the probability amplitude for the particle (Universe) being maximally localized around the position $\zeta$, is nondiverging for all $\zeta$, as soon as we take the condition $C_1=0$. We stress that the canonical wave function (the function (\ref{solcan})) is diverging at the classical singularity $x=0$.

To get a better feeling with our quantum Universe we construct and examine the motion of wave packets. Let's now construct states peaked at late times
\be\label{wp} 
\Psi(\zeta,t)=\int_0^\infty d\epsilon g(\epsilon)\Psi_\epsilon(\zeta)e^{i\epsilon t},
\ee
where we have defined the dimensionless time $t=\sqrt{24\pi\kappa}\phi$. In the following we take $g(\epsilon)$ to be a Gaussian distribution peaked at some $\epsilon^\ast\ll1$, which corresponds to be peaked at energy much less then the Plank energy $1/l_P$ (we recall that $\epsilon\sim\mathcal{O}(\overline\epsilon\ l_P)$, where $\overline\epsilon$ have dimension $1$ in energy). The analytic computation of the integral (\ref{wp}) for the wave function (\ref{quapos}) is impossible to perform. So, in order to describe the motion of wave packets we have to evaluate (\ref{wp}) numerically.
\begin{figure}
\includegraphics[height=1.8in]{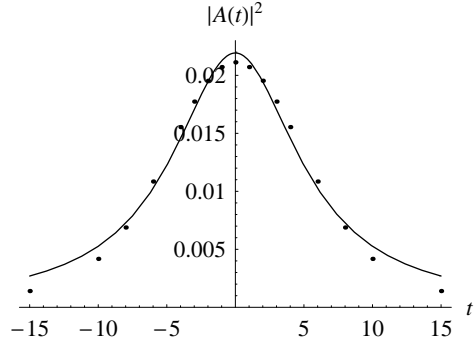}
\caption{The points represent the result of the numerical integration and are fitted by a Lorentzian $L(t)=0.692/(t^2+31.564)$ heaving width, at the inflection point, $3.243$.} 
\end{figure}

At first, we want to analyze the most interesting region, i.e. where $\zeta\simeq0$, which corresponds to the purely quantum region, where the physical volume is Planckian. In fact, if we put $\beta\sim\mathcal{O}(l_P^6)$, the minimal uncertainty in position is of order of the Planckian volume. The ``quasiposition wave function'' (\ref{quapos}) can be expanded in order to give the probability density around $\zeta\simeq0$: $\vert\Psi(\zeta,t)\vert^2\simeq\vert A(t)\vert^2+\zeta^2\vert B(t)\vert^2$. Therefore, starting with a state peaked at some $\epsilon^\ast\ll1$, the probability density of finding the Universe ``around the Planckian region'' is $\vert A(t)\vert^2$, where $A(t)$ reads
\be\label{At}
A(t)=2C_2\int_0^\infty d\epsilon\f{(1+2\sqrt{1-\epsilon^2})e^{-\f{(\epsilon-\epsilon^\ast)^2}{2\sigma^2}+i\epsilon t}}{\sqrt{1-\epsilon^2}\left(3-\epsilon^2+3\sqrt{1-\epsilon^2}\right)} .
\ee
We evaluate the above integral numerically for $\epsilon^\ast=10^{-3}$, $\sigma^2=1/20$ and we take the constant $2C_2=1$. The probability density $\vert A(t)\vert^2$ is very well approximated by a Lorentzian function (see Fig. 1). As we can see form the picture, this curve is peaked around $t=0$. This value corresponds to the classical time for which $x(t)=x_0$ (in (\ref{At}) we consider $t_0=0$). Thus, for $x_0\sim\mathcal{O}(l_P^3)$, the probability density to find the Universe in a Planckian volume is peaked around the corresponding classical time. As a matter of fact this probability density vanishes for $t\rightarrow-\infty$, where the classical singularity appears. This is the meaning when we claim that the classical cosmological singularity is solved by this model. 
\begin{figure}
\includegraphics[height=1.8in]{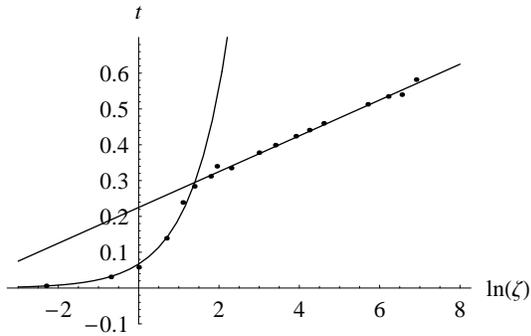}
\caption{The peaks of the probability density $\vert\Psi(\zeta,t)\vert^2$ are plotted as functions of $t$ and $\ln(\zeta)$. The points (resulting from numerical computation) are fitted by a logarithm $0.050\ln(\zeta)+0.225$ for $\zeta\geq4$ and by a power law $0.067\zeta^{1.060}$ for $\zeta\in[0,4]$.} 
\end{figure}

In order to describe the motion of the wave packet we evaluate $\vert\Psi(\zeta,t)\vert^2$ from the integral (\ref{wp}) of the wave function (\ref{quapos}). As before, we consider a wave packet initially peaked at late times and let it evolve numerically ``backward in time''. We use the same parameters for the integration performed above. The result of the integration is that the probability density, at different fixed values of $\zeta$, is very well approximated by a Lorentzian function yet. The width of this function remains, actually, the same as the states evolves from large $\zeta$ ($10^3$) to $\zeta=0$. For all fixed $t$ the probability density is well-fixed by a Lorentzian function and the width of these function, also in this case, remain almost the same during the evolution. These states are sharply peaked for $\zeta\sim\mathcal{O}(1)$ (which in our units correspond to $\zeta\sim\mathcal{O}(\sqrt\beta)\sim\mathcal{O}(l_P^3)$). The peaks of Lorentzian functions, at different $\zeta$ values, move along the classically expanding trajectory (\ref{clastra}) for values of $\zeta$ larger then $\sim4$. Near the Planckian region, i.e. when $\zeta\in[0,4]$, we observe a modification of the trajectory of the peaks. In fact they follow a power-law up to $\zeta=0$, reached in a finite time interval and ``escape'' from the classical trajectory toward the classical singularity (see Fig. 2). The peaks of the Lorentzian at fixed time $t$, evolves very slowly remaining close to the Planckian region. Such behavior outlines that the Universe has a stationary approach to the cutoff volume, accordingly to the behavior in Fig. 2.  

This peculiar behavior of our quantum Universe is different from other approaches to the same problem. In fact, recently, it was shown how the classical Big-Bang is replaced by a Big-Bounce in the framework of Loop Quantum Cosmology (LQC) \cite{APS}. Intuitively, one can expect that the bounce and so the consequently repulsive features of the gravitational field in the Planck regime are consequences of a Planckian cut-off length. But this is not the case. As matter of fact, we can observe from Fig. 2 that there is not a bounce for our quantum Universe. The main differences between the two approaches resides in the quantum modification of the classical trajectory. In fact, in the LQC framework we observe a ``quantum bridge'' between the expanding and contracting Universes; in our approach, contrarily, the probability density of finding the Universe reaches the Planckian region in a stationary way.

\vspace{0.2cm}

Let us now reassume the main differences between the Wheeler-DeWitt and the Generalized Uncertainty Principle approaches to the flat FRW Universe filled by a massless scalar field. The first distinction reside on a probabilistic level, i.e. on the diverging of the wave function at the classical singularity. The second, and more relevant, difference concerns the dynamics of the wave packets toward the singularity. In particular, we have:

i) The WDW wave function (\ref{solcan}) is diverging at the classical singularity $a=0$. Therefore, the corresponding probability defined above is diverging, i.e.
\be    
P_{\delta}\equiv\int_0^\delta\vert\Psi_{WDW}(a,t)\vert^2da=\infty.
\ee
On the other hand, the GUP wave function (\ref{quapos}) is nondiverging for all the ``quasiposition'' $\zeta$, as soon as the condition $C_1=0$ is taken. In this respect we obtain 
\be    
P_{\delta}\equiv\int_0^\delta\vert\Psi_{GUP}(a,t)\vert^2da<\infty.
\ee
Therefore, already at this level, we can claim to have a no collapse behavior for the quantum GUP Universe. Of course this is not the case for the WDW theory. 

ii) The semi-classical wave packets, in the WDW scheme, fall into the Big-Bang singularity. More precisely, it is possible to construct a wave packet which is peaked at late time, i.e. far from the Planckian region. Then, in the backward evolution toward the singularity, the wave packet continues to be peaked on the classical trajectory (\ref{clastra}) for all the ``times'' and therefore can not escape from the classical singularity. In this sense, the WDW approach does not resolve the singularity problem. On the other hand, the GUP wave packets do not fall into the singularity. In particular, at a given ``time'', they escape from the classical trajectory and the Universe exhibit a stationary behavior in approaching the Planckian volume. This way, the classical singularity is solved by our model.  

\section{VI. The Taub model}

The Taub model is a particular case of the Bianchi IX model. This model is (together with Bianchi VIII) the most general homogeneous cosmological model and its line element reads\footnote{From now on we work in $\hbar=c=16\pi G=1$ units.}, in the Misner parametrization \cite{Mis69}, 
\be
ds^2=N^2dt^2-e^{2\alpha}\left(e^{2\gamma}\right)_{ij}\omega^i\otimes\omega^j,
\ee
where $N=N(t)$ is the lapse function and the left invariant 1-forms $\omega^i=\omega^i_adx^a$ satisfy the Maurer-Cartan equations $2d\omega^i=\epsilon^i_{jk}\omega^j\wedge\omega^k$. The variable $\alpha=\alpha(t)$ describes the isotropic expansion of the Universe and $\gamma_{ij}=\gamma_{ij}(t)$ is a traceless symmetric matrix $\gamma_{ij}=diag\left(\gamma_++\sqrt3\gamma_-,\gamma_+-\sqrt3\gamma_-,-2\gamma_+\right)$ which determines the shape change (the anisotropy) {\it via} $\gamma_\pm$. Since the determinant of the 3-metric is given by $h=\det e^{\alpha+\gamma_{ij}}=e^{3\alpha}$, it is easy to recognize that the classical singularity appears for $\alpha\rightarrow-\infty$. 

Performing the usual Legendre transformations, we obtain the Hamiltonian constraint for this model. As well-known \cite{Mis69,CGM} the dynamics of the Universe, toward the singularity, is described by the motion of a two-dimensional particle (the two physical degree of freedom of the gravitational field) in a dynamically-closed domain. In the Misner picture, such a domain depends on the time variable $\alpha$ and therefore to overcame this difficulty, the so-called Misner-Chitr$\acute e$-like variables \cite{Chi} are introduced. In such a scheme the dynamically-allowed domain becomes independent on the new time variable. 

The next step is to perform the ADM reduction of the dynamics. This scheme relies on the idea to solve the classical constraint, with respect to a given momenta, before implementing some quantization algorithm. In this way, we will obtain an effective Hamiltonian which will depend only on the physical degrees of freedom of the system. Moreover, it is possible to choose the so-called Poincar$\acute e$ representation in the complex upper half-plane \cite{KM97}, in which the ADM ``constraint'' becomes
\be\label{huv}
-p_\tau\equiv H_{ADM}^{IX}=v\sqrt{p_u^2+p_v^2},
\ee
being $\tau$ the new time variable and $u,v$ related to the anisotropies variables $\gamma_\pm$.

The Taub model is nothing but the Bianchi IX model in the $\gamma_-=0$ case \cite{RS}. The dynamics of this Universe is equivalent to the motion of a particle in a one-dimensional closed domain. Such a domain corresponds to take only one of the three equivalent potential walls of the Bianchi IX model. It is no difficult to see that this particular case appears for $u=-1/2$ and therefore the ADM Hamiltonian (\ref{huv}) rewrite 
\be\label{hv}
H_{ADM}^T=vp_v,
\ee
being $v\in[1/2,\infty)$. The above Hamiltonian (\ref{hv}) can be further simplified defying a new variable $x=\ln v$ and becomes
\be\label{ht}
H_{ADM}^T=p_x\equiv p,
\ee
which will be the starting point of the upcoming analysis. Let us stress that the classical singularity now appears for $\tau\rightarrow\infty$. 

\section{VII. Quantum Taub dynamics in the WDW scheme}

In this section we focus our attention on the canonical quantum features of the Taub Universe, described by the Hamiltonian (\ref{ht}) with the boundary condition $x\in[x_0\equiv\ln(1/2),\infty)$. In particular, without discuss the computation details, we construct and analyze the motion of suitable wave packets in the $(\tau,x)$-plane. The result is plotted in Fig. 3 and the physical meaning of the configuration variable $x$ is clarified by the relation
\be\label{anix}
\gamma_+=\f{e^{\tau-x}}{\sqrt3}\left(e^{2x}-\f34\right).
\ee
\begin{figure}
\includegraphics[height=1.8in]{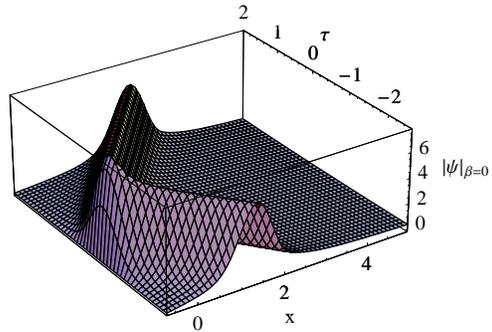} 
\caption{The evolution of the wave packets $\vert\Psi(\tau,x)\vert$ in the canonical case, i.e. $\beta=0$. The $x$ variable is in the $[x_0,5]$-interval.} 
\end{figure}

As we can see from the picture, the wave packets follow the classical trajectories (for more details see \cite{BM07b}). The probability amplitude to find the particle (Universe) is packed around these trajectories. In this respect no privileged regions arise, namely no dominant probability peaks appear in the ($\tau,x$)-plane. As matter of fact, the ``incoming'' Universe ($\tau<0$) bounce at the potential wall at $x=x_0$ and then fall toward the classical singularity ($\tau\rightarrow\infty$). Therefore, as well-known, the Wheeler-DeWitt formalism is not able to get light on the necessary quantum resolution of the classical cosmological singularity. As we will see in the next section, this picture is radically changed in the GUP framework.

\section{VIII. Quantum Taub dynamics in the GUP scheme}

Let us now analyze the quantum evolution of the Taub Universe in the deformed Heisenberg algebra formalism \cite{BM07b}. Namely, we perform a generalized quantization of this model based on the GUP approach. Let us stress that, from the ADM reduction of the dynamics, the variable $\tau$ is regarded as a time coordinate and therefore the conjugate couple ($\tau,p_\tau$) will be treated in a canonical way. This way, considering the Hamiltonian (\ref{ht}), we deal with a Sch\"odinger-like equation
\be\label{eqschtaub}
i\p_\tau\Psi(\tau,p)=\hat H_{ADM}^T\Psi(\tau,p).
\ee
As we have explained before, in the GUP approach, we have lost all informations on the position itself. Therefore, the boundary condition have to be imposed on the ``quasiposition wave function'' (\ref{qwf}), in the sense that $\psi(\zeta_0)=0$ (being $\zeta_0=\langle\psi^{ml}_{\zeta}\vert x_0\vert\psi^{ml}_{\zeta}\rangle$ in agreement with the previous discussion). The solution of the eigenvalue problem resulting from (\ref{eqschtaub}), is the Dirac $\delta$-distribution $\psi_k(p)=\delta(p^2-k^2)$ and therefore the ``quasiposition wave function'' (\ref{qwf}) reads 
\begin{multline} 
\psi_k(\zeta)=\f1{k(1+\beta k^2)^{3/2}}\left[A\exp\left(i\f{\zeta}{\sqrt{\beta}} \tan^{-1}(\sqrt{\beta}k)\right)\right.+\\+\left.B\exp\left(-i\f{\zeta}{\sqrt{\beta}} \tan^{-1}(\sqrt{\beta}k)\right)\right],
\end{multline}
where $A$ and $B$ are integration constants. In this way, the boundary condition $\psi(\zeta_0)=0$ can be easily imposed and fixes one constant, giving us the final form for the ``quasiposition'' eigenfunctions
\begin{multline}\label{ef}
\psi_k(\zeta)=\f A{k(1+\beta k^2)^{3/2}}\left[\exp\left(i\f{\zeta}{\sqrt{\beta}} \tan^{-1}(\sqrt{\beta}k)\right)+\right.\\-\left.\exp\left(i\f{(2\zeta_0-\zeta)}{\sqrt{\beta}} \tan^{-1}(\sqrt{\beta}k)\right)\right].
\end{multline}

Let us construct and examine the evolution of wave packets. The analysis of dynamics of such wave packets allow us to give a precise description of the evolution of the Taub Universe. They are superposition of the eigenfunctions, i.e.
\be\label{wapa}
\Psi(\tau,\zeta)=\int_0^\infty dk A(k)\psi_k(\zeta)e^{-ik\tau}.
\ee
In the following we will take $A(k)$ as a Gaussian-like function
\be\label{gau}
A(k)=k(1+\beta k^2)^{3/2}e^{-\f{(k-k_0)^2}{2\sigma^2}}
\ee
in order to simplify the explicit expression of the wave packets. The computation of (\ref{wapa}) for the eigenfunctions (\ref{ef}) is performed in a numerical way and the parameters are chosen as follows: $k_0=1$ and $\sigma=4$.

As we said the parameter $\beta$, i.e. the presence of a nonzero minimal uncertainty in the configuration variable ($\Delta x_{min}=\sqrt\beta$), is responsible for the GUP effects on the dynamics. In fact, in the $\beta=0$ limit, the eigenfunctions (\ref{ef}) reduce to ordinary plane waves and the quasiposition $\zeta\rightarrow x$, i.e. we recover the WDW scheme. 
\begin{figure}
\includegraphics[height=1.8in]{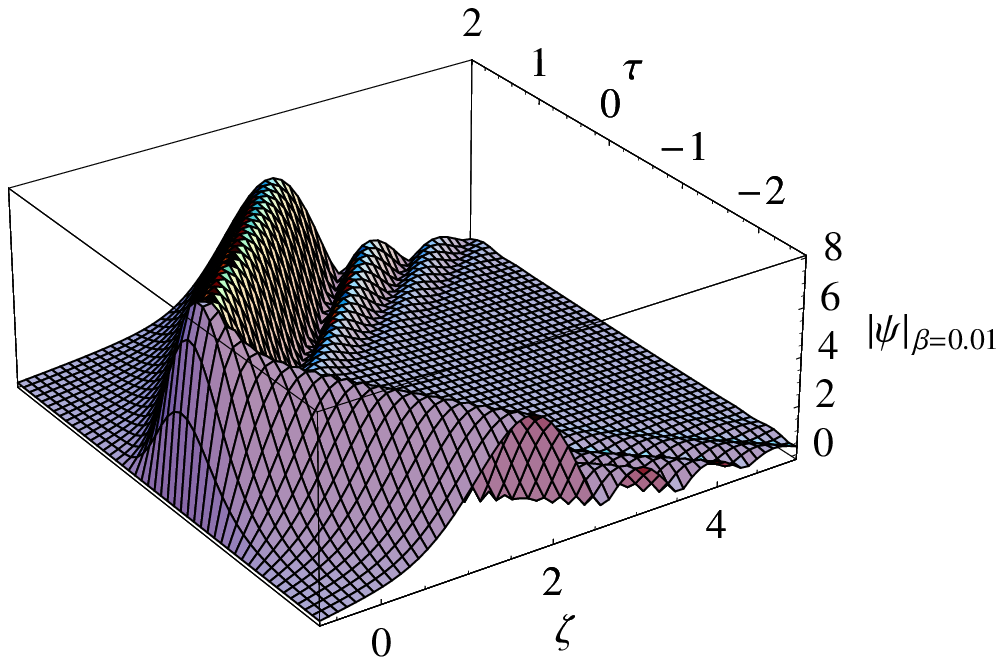} \includegraphics[height=1.8in]{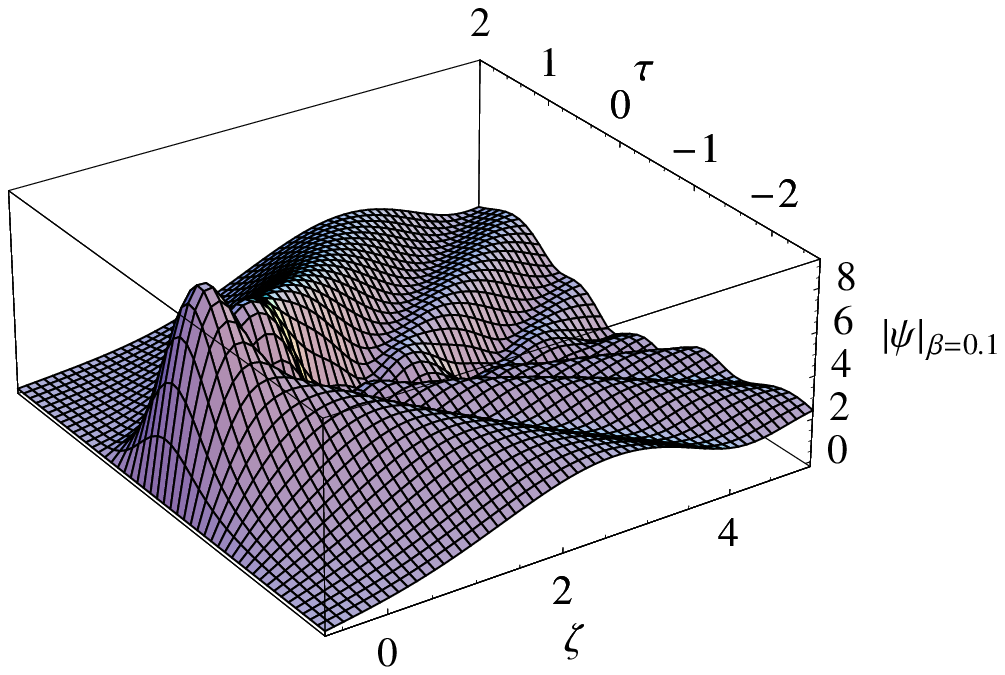} \includegraphics[height=1.8in]{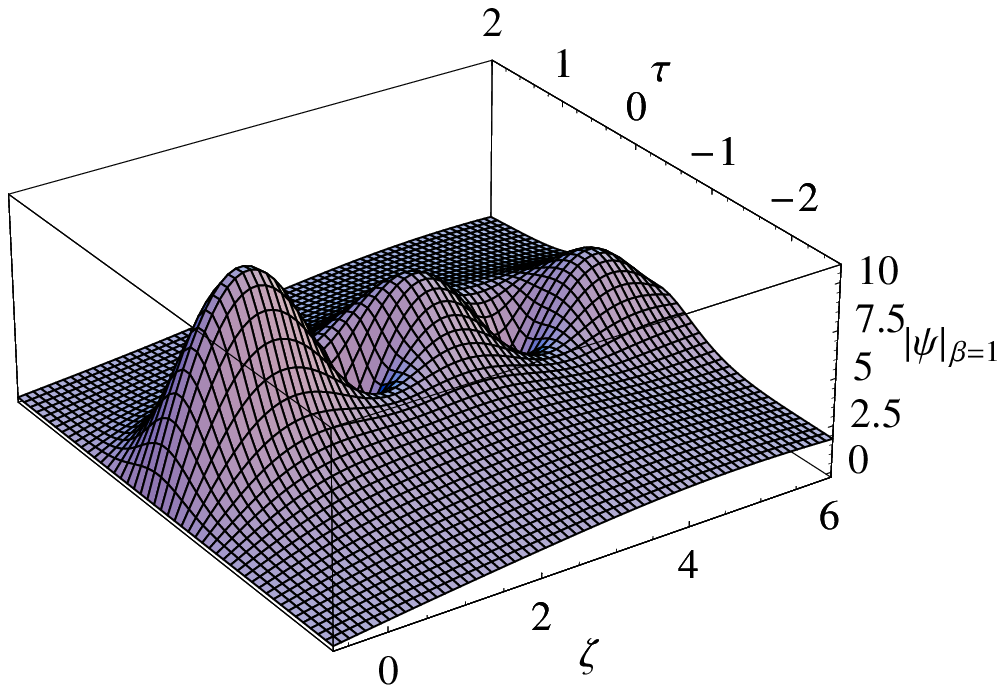}
\caption{The evolution of the wave packets $\vert\Psi(\tau,\zeta)\vert$ in the GUP framework. The graphics are for $\beta=0.01$, $\beta=0.1$ and $\beta=1$ respectively. For smaller $\beta$ the canonical case is recovered.} 
\end{figure}

Therefore, in order to comprehend the alterations induced by the deformed Heisenberg algebra on the canonical Universe dynamics, we have to analyze different $\beta$-regions. In fact, when the ``deformation'' parameter $\beta$ become more and more important, i.e. when we are at some scale which allows us to appreciate the GUP effects, the evolution of the wave packets is different from the canonical case. In particular, we can distinguish between three different $\beta$-regimes:
\begin{itemize}
	\item Let us first consider the ($\beta\sim\mathcal O(10^{-2})$)-region. In this regime the wave packets begin to spread and a constructive and destructive interference between the incoming and outgoing wave appears. The probability amplitude to find the Universe is still peaked around the classical trajectory, but ``not so much'' as in the canonical case.
\end{itemize}
 
\begin{itemize}
	\item When this parameter becomes more influent, i.e. $\beta\sim\mathcal O(10^{-1})$, we can no more distinguish an incoming or outgoing wave packet. At this level is meaningless to speak about a wave packet which follows the classical trajectory. Moreover, the probability amplitude to find the Universe is, in some sense, pecked in a specified region in the ($\tau,\zeta$)-plane, i.e. for $\zeta\simeq0$.
\end{itemize}
 
\begin{itemize}
	\item As last step, for $\beta\sim\mathcal O(1)$, a dominant probability peak ``near'' the potential wall appears. In this $\beta$-region, there are also other small peaks for growing values of $\zeta$, but they were widely suppressed for bigger $\beta$. In this case, the motion of wave packets show a stationary behavior, i.e. these are independent on $\tau$.
\end{itemize}
Following this picture we are able to learn the GUP modifications to the WDW wave packets evolution. In fact, considering a sort of dynamics in the ``deformation'' parameter $\beta$, from small to ``big'' values of $\beta$, we can see how the wave packets ``escape'' from the classical trajectories and approach a stationary state close to the potential wall. All this picture is plotted in Fig. 4. 

From this point of view, the classical singularity ($\tau\rightarrow\infty$) is widely probabilistically suppressed, because the probability to find the Universe is peaked just around the potential wall. Another feature to be considered, is that the large anisotropy states are not privileged. In fact, the most probable states, as we can see from the picture, are those for $\zeta\simeq0$, i.e. from equation (\ref{anix}) we obtain $\vert\gamma_+\vert\simeq e^\tau/10$. Therefore, with respect to predict an isotropic Universe, the GUP wave packets exhibit a better behavior with respect those in the WDW theory.

\section{IX. Concluding Remarks}

The effects of a modified Heisenberg algebra, which reproduces a GUP as appeared in studies on String Theory \cite{String}, on the Big-Bang singularity and on the Taub model are showed. 

In the case of the flat FRW Universe, the evolution is performed with respect to the scalar field taken as an emergent time and the the model appears to be singularity-free. Furthermore, suitable wave packets were constructed and their dynamics toward the classical singularity analyzed. As matter of fact, such a Universe show a stationary feature toward the Planckian region and no evidence for a Big-Bounce seems to come out.

The dynamics of Taub model, on the other hand, was investigated in terms of an internal variable, related to the Universe isotropic volume. Also in this case the Universe exhibit a singularity-free behavior. As matter of fact, the wave packets stop following the classical trajectories toward the singularity and a dominant peak (near the potential wall region), in the probability amplitude to find the Universe, appears. Moreover, the large anisotropy states, i.e. those for $|\gamma_+|\gg1$ ($|x|\gg1$), are probabilistically suppressed. 
  
\bibliographystyle{aipproc}   

\begin{thebibliography}{99}

\bibitem{String}D.J.Gross and P.F.Mendle, Nucl.Phys.B {\bf 303} (1988) 407; D.J.Gross, Phys.Rev.Lett. {\bf 60} (1988); 1229 D.Amati, M.Ciafaloni and G.Veneziano, Phys.Lett.B {\bf 216} (1989) 41; E.Witten, Phys.Today {\bf 49} (1997) 24; G.Amelino-Camelia et al., Mod.Phys.Lett.A {\bf 12} (1997) 2029.

\bibitem{Mag}M.Maggiore, Phys.Lett.B {\bf 304} (1993) 65.

\bibitem{Sny}H.S.Snyder, Phys.Rev. {\bf 71} (1947) 38. 

\bibitem{GUP1}A.Kempf and L.Lorenz, Phys.Rev.D {\bf 74} (2006) 103517; A.Ashoorioon, A.Kempf and R.B.Mann, Phys.Rev.D {\bf 71} (2005) 023503; L.N.Chang et al., Phys.Rev.D {\bf 65} (2002) 125027; F.Brau and F.Buisseret, Phys.Rev.D {\bf 74} (2006) 036002; J.Y.Bang and M.S.Berger, Phys.Rev.D {\bf 74} (2006) 125012; M.M.Stetsko, Phys.Rev.A {\bf 74} (2006), 062105; G.Amelino-Camelia et al., CQG {\bf 23} (2006) 2585. 

\bibitem{Vakili}B.Vakili and H.R.Sepangi, Phys.Lett.B {\bf 651} (2007) 79; A.Bina et al., IJTP (in press), arXiv:0709.3623. 

\bibitem{BM07a}M.V.Battisti and G.Montani, Phys.Lett.B (in press), arXiv:gr-qc/0703025. 

\bibitem{BM07b}M.V.Battisti and G.Montani, {\it Quantum Dynamics of the Taub Universe in a Generalized Uncerainty Principle framework}, arXiv:0707.2726.

\bibitem{Ish}W.F.Blyth and C.J.Isham, Phys.Rev.D {\bf 11} (1975) 768.

\bibitem{APS}A.Ashtekar et al., Phys.Rev.Lett. {\bf 96} (2006) 141301; A.Ashtekar et al., Phys.Rev.D {\bf 73} (2006) 124038.

\bibitem{Kem}A.Kempf, G.Mangano and R.B.Mann, Phys.Rev.D {\bf 52} (1995) 1108; A.Kempf, J.Math.Phys. {\bf 38} (1997) 1347.

\bibitem{Got}M.J.Gotay and J.Demaret, Phys.Rev.D {\bf 28} (1983) 2402.

\bibitem{DW}B.DeWitt, Phys.Rev. {\bf 160} (1967) 1113.

\bibitem{Mis69}C.Misner, Phys.Rev.Lett., {\bf 22} (1969) 1071.

\bibitem{CGM}G.P.Imponente and G.Montani, Phys.Rev.D, {\bf 63} (2001) 103501; R.Benini and G.Montani, Phys.Rev.D, {\bf 70} (2004) 103527.

\bibitem{Chi}D.M.Chitr$\acute e$, {\it PhD Thesis}, University of Maryland (1972).

\bibitem{KM97}A.A.Kirillov and G.Montani, Phys.Rev.D, {\bf 56} (1997) 6225.

\bibitem{RS}M.P.Ryan and L.C.Shapley, {\it Homogeneous Relativistic Cosmologies} (PUP, Princeton, 1975).

\end{thebibliography}

\end{document}